
\magnification=1200
\parindent 1.0truecm
\baselineskip=16pt
\rm
\null


\footline={\hfil}
\vglue 0.8truecm
\rightline{\bf DFUPG-36-91-rev}
\rightline{\sl November 1991 (revised February 1992)}

\vglue 2.0truecm
\centerline{\bf OLD PROBLEMS IN LOW-ENERGY $K N$ PHYSICS }
\centerline{\bf AND PERSPECTIVES OPENING UP AT NEW MACHINES }

\vglue 1.0truecm
\centerline{\sl Paolo M. Gensini }
\centerline{\sl Dip. di Fisica dell\'\ Universit\`a di Perugia, Italy, and }
\centerline{\sl I.N.F.N., Sezione di Perugia, Italy }
\centerline{ and }
\centerline{\sl Galileo Violini }
\centerline{\sl Centro Internacional de F\'\i sica, Santa Fe de Bogot\'a,
Colombia, }
\centerline{\sl Dip. di Fisica dell\'\ Universit\`a della Calabria, Rende
(Cosenza), Italy, and }
\centerline{\sl I.N.F.N., Gruppo Collegato di Cosenza, Italy. }

\vglue 1.7truecm
\centerline{\bf ABSTRACT }
\vglue 0.3truecm

\par We survey the open problems in low--energy $KN$ physics, in the
perspective of using the new machines presently planned, such as the
$\phi$--factory DA$\Phi$NE at I.N.F.N. Nat. Labs. in Frascati
(as a source of {\sl tagged}, low--energy kaons) and the
KAON Factory at TRIUMF (as a producer
of intense, intermediate--energy kaon beams).

\vglue 2.0truecm
\centerline{ Contribution presented at the }
\centerline{\sl International Symposium on Hypernuclear and Strange Particle
Physics }
\centerline{\sl Shimoda, Japan, December 9--12, 1991 }
\pageno=0
\vfill
\eject

\footline={\hss\tenrm\folio\hss}

\vglue 0.5truecm
\noindent{\bf 1. INTRODUCTION.}
\vglue 0.3truecm

\par The aim of this contribution is to review the
state--of--the--art of both our experimental and theoretical knowledge of
the low--energy $K N $ physics, and to stress the importance of
better information on several of
its parameters, notably the $K N Y$ coupling constants and the $ K N$
sigma terms.

\par Our main motivation stems from the possibilities opening up in
the next few years, when new machines such as DA$\Phi$NE and KAON will
become operational. They will make possible again to perform systematic
studies of $K N$ interactions, whose importance was stressed in the
past$^1$, in connection with the closing down of low--energy machines and/or
kaon beam lines (particularly with the demolition of NIMROD and the closing
of CERN beam lines); this left the remaining beam lines, at BNL and KEK,
to carry all the load of the recent research with strange particles.

\par Now let us shortly describe the main, relevant features of
DA$\Phi$NE and KAON, in order to elucidate their complementarity for such a
systematic investigation.

\par DA$\Phi$NE (``Double Annular $\Phi$--factory for Nice Experiments'') is
the $\phi$--factory, due to replace Adone at the Frascati National
Laboratories of I.N.F.N., and has already been approved as part of the
Institution's five--year plan.
Because of its high commissioning luminosity$^2$,
its {\sl two} interaction regions will be sources of $\simeq 436\ K^\pm$
s$^{-1}$, at a momentum of 126.9 MeV/c, with a momentum resolution of
$\simeq 1.1 \times 10^{-2}$, as well as of $\simeq 303\ K_L$ s$^{-1}$, at a
momentum of 110.1 MeV/c, with the slightly worse resolution of $\simeq 1.5
\times 10^{-2}$.

\par This will make possible to perform $K^\pm N$ experiments from
the elastic threshold up to about 125 MeV/c and $K_L N$ experiments
at 110 MeV/c
with practically no background, since those from both $\pi^\pm$'s
and leptons are easy to eliminate at a trigger level:
because the $\pi^\pm$'s come {\sl almost all} from events with three or more
final particles, the former can be suppressed by momentum
and acollinearity cuts, while
the latter are completely eliminated by a momentum cut, which also eliminates
collinear pions from $\phi \rightarrow \pi^+ \pi^-$.

\par The KAON (``Kaons, Antiprotons, Other hadrons and Neutrinos'') Factory
design is an accelerator complex, recently approved by the Canadian Federal
Government, which will use the existing TRIUMF $H^-$ cyclotron as an
injector$^3$. Six charged--kaon channels are foreseen with momenta ranging
 from 0.4 to 21 GeV/c, and the intensities of the beams
in the lowest momentum range
below 800 MeV/c go from a few to about a hundred millions $K^-$ s$^{-1}$,
$K^+$ beams being about twice more intense$^4$.

\par Decays in flight limit the purity of fixed--target--machine kaon beams:
to perform measurements at momenta below 400 MeV/c, one has to use
moderators, which at the same time decrease the kaon intensity and increase
the beam contamination at the final target. This contamination requires
identification of the incoming particles either via additional TOF or
\v Cerenkov apparatuses: thus the advantage provided by the high intensities
has to be balanced against the higher costs of the experiments. This situation
contrasts that at the $\phi$ factories, where the intensities are lower by
orders of magnitude, but contaminations can be controlled without additional
setups.

\par The coming in operation of KAON and DA$\Phi$NE will improve significantly
our possibility of making low--energy $K N$ experiments, both increasing
statistics and covering in detail regions barely explored in the past:
this will allow new, much more accurate analyses. Therefore
in this contribution we want to discuss
which phenomenological aspects of $K N$ physics may benefit from the
situation. For this purpose we shall briefly review the present status of
the phenomenology and mention a number of problems which are still open.

\vglue 0.6truecm
\noindent{\bf 2. THE STATUS OF PRESENT INFORMATION ON $KN$ INTERACTIONS AT
LOW MOMENTA. }
\vglue 0.3truecm

\par The low--energy interactions for both $K^+ N$ and $K^- N$ are
described in terms of phase shifts. The amplitudes for the two
sectors ($S = +1$ and $S = -1$) are related by analyticity and crossing;
additionally, the regeneration amplitude is a linear combination of the
two. In the following we shall give a picture of the situation of the
phase shifts for each sector separately. It has to be observed that this
information comes basically from the charged--kaon induced reactions:
although it has occasionally been suggested to analyze simultaneously
these data and the regeneration ones$^5$, the statistical weight of the
latter has been insufficient to provide more than a mere consistency check$^6$.

\par Let us start with the $S = +1$ sector. Most $K^+N$ data come
 from experiments performed more than ten years ago, excepting some
higher--momentum polarization data coming from KEK. The
low--energy region is described by $S-$ and $P-$waves for both $I = 0, 1$
isospin channels: while $K^+p$ scattering is pure $I = 1$, with a repulsive
$S$--wave and a weak $P$--wave which starts being felt only in the
higher--momentum range, $K^+n$ scattering (whose extraction from $K^+d$
requires a treatment of the three--body problem for the $Knp$
system) contains also an $I = 0$ amplitude, characterized by a small $S-$wave
scattering length and a weak $P-$wave (as in the $I = 1$ channel).

\par The general behavior of the low--energy amplitudes is thus rather well
understood$^7$. Interactions of $K^+$'s with the nuclear constituents
correspond to a mean--free path of about 7 fm, so that they are sensitive
probes of nuclear matter; in the last few years much debate has been
given to the comparison of $K^+$ scattering on
nucleons with that on nuclei (in particular on carbon), which led to the
suggestive hypothesis of a possible swelling of nucleons in nuclei$^8$. The
present uncertainty on the relevant data does not allow sharp predictions,
so that more precise information is necessary.

\par Turning now to the $S = -1$ sector, the most typical
features of low--energy $\bar KN$ scattering are its
large unphysical regions, which start at the $\pi \Lambda$ threshold for
$I = 1$ and at the $\pi \Sigma$ one for $I = 0$, and extend to the elastic
$\bar KN$ one. The importance of these regions is enhanced by the presence
of both the $P$--wave $\Sigma(1385)$ and the $S$--wave $\Lambda(1405)$
resonances: the scattering amplitudes there have to be analytic continuations
of the corresponding ones in the low--energy $\bar KN$ channels$^9$.

\par The low--energy amplitudes are described by a partial--wave formalism
which has to include: a) couplings between {\sl all} the open channels,
b) analyticity in the energy plane,
and c) unitarity in all channels considered.
The most common
one is the K--matrix formalism$^{10-12}$, where one writes the
partial--wave $T$--matrix $\hat T_I^l$ as
$$
(\hat T_I^l)^{-1} = (\hat K_I^l)^{-1} - i \hat Q^{(2 l + 1)} \ .
\eqno (1)
$$
Here $\hat K_I^l$ is a real Hermitian matrix, and $\hat Q$ is the diagonal
matrix of the c.m.s. momenta $q_i$ (or a modification of the latter, analytic
in the cut $s$--plane and having the same behavior at the thresholds). Several
parameters are usually required to describe the low--energy region above {\sl
and} below threshold, leaving much too large a margin for ambiguities, given
the quality of presently available data, unless
the parameters are otherwise constrained by suitable theoretical arguments.

\par A little accounting shows that, even neglecting the weakly coupled $\pi
\pi \Lambda$ channel, the formalism requires, for each coefficient of a power
series in $s$ for either $\hat K_I^l$ or its inverse $\hat M_I^l$, three
parameters for $I = 0$ and six for $I = 1$, or a total of 36 parameters
to describe all channels up to order $q^3$ with $S$-- and $P$--waves.

\par Note that the charge states of the same system have been counted as {\sl
one} channel: though the differences in threshold energies could be considered
purely electromagnetic effects, the situation is here more complex than in the
$\pi N$ case, even in the physical region.
We have indeed $\omega_{\bar K^0n} - \omega_{K^-p} = + 8.13$ MeV,
corresponding to a $p_L(K^-) = 90$ MeV/c, versus $\omega_{\pi^0n} -
\omega_{\pi^-p} = - 3.79$ MeV: thus $\bar K^0n$ threshold effects (cusps
and/or dips) will be observable in the momentum region covered by the
$\phi$--factory kaons, and could provide valuable information
on the interplay between
electromagnetic and strong interactions in the K--matrix representation
of the $\bar KN$ amplitudes.

\par Presently available low--energy data consist essentially of {\sl
integrated} cross sections, with extremely scarce angular distributions and
polarizations, coming in only at higher energies; as mentioned above,
this, and the limited
statistics, do not allow to
determine uniquely the parameters, and additional constraints are usually
introduced.

\par As a typical example of this procedure one can mention the use of
fixed--$t$ dispersion relations$^{10}$ (for
symbols and conventions, see Ref. 9)
$$
D^+(\omega) = D^+(0) + \sum_Y {{\omega R_Y g_Y^2} \over {\omega_Y
(\omega - \omega_Y)}} + {\omega \over \pi} \int_{\omega_{thr}}^\infty
[{A^+(\omega^{\prime}) \over {\omega^{\prime} (\omega^{\prime} - \omega)}}
- {A^-(\omega^{\prime}) \over {\omega^{\prime} (\omega^{\prime} + \omega)}}]
d\omega^{\prime}  ,
\eqno (2)
$$
connecting the threshold amplitudes and the integrals over the unphysical
region, entering the last term (given by fits to the low--energy data), to
data on real and imaginary parts of the amplitudes at higher energies.

\par Good knowledge of the real parts is essential in making such constraints
tight, calling in turn for high--quality measurements of elastic cross
sections in the Coulomb--nuclear interference region:
the accuracy of available data has till now limited the use of
these constraints to $t = 0$, where the imaginary part of the amplitude
is provided by the optical theorem.
Drastic improvements over the quality of the data used for amplitude
analyses$^7$ (both $d\sigma /d\Omega$ and polarization $P$) are
required to extend them also to $t \not= 0$.

\par In this context one has to remark that the validity of dispersion
relations is related to that of microcausality at short distances; the use
of dispersive constraints in data analysis may then weaken the significance
of the tests on the above validity. Therefore the availability of new,
accurate, low--energy data should be regarded as an essential contribution
to an unbiased understanding of the basics of fundamental interactions.

\par A general feature of all the parametrizations is a strongly
absorptive interaction in both isospin channels. However, the detailed
structure of the amplitudes, particularly in the unphysical region, is far
 from being determined, mainly due to the low statistics of the experiments
performed up to now on inelastic production of these low--mass
systems$^{13}$.
\vfill
\eject

\noindent{\bf 3. THE STILL UNANSWERED PROBLEMS OF LOW--ENERGY $KN$ PHY-SICS. }
\vglue 0.3truecm

\par The uncertainties discussed in the previous section, on
the determination of the low--energy $\bar KN$
and $KN$ amplitudes, reflect on uncertainties in the evaluation of a number
of quantities, which are fundamental to our understanding of
strong--interaction physics in the strange--particle sector. In particular
we are referring to the $KNY$ coupling constants, to the $\sigma$--terms,
and to the still open puzzle posed by the energy shift and the width of
the X--ray lines of kaonic hydrogen.

\par Coupling constants were once considered fundamental, or studied
as tests of $SU(3)_f$ symmetry; they have regained status as sensitive
tests of the structure of the {\sl explicit} chiral--symmetry breaking
of QCD$^{14}$, via their deviations $\Delta_{KNY}(0)$ from
Gold\-ber\-ger--Trei\-man
relations, defined by
$$
\sqrt{1/2} (M_N + M_Y) G_A^{YN}(0) = f_K\ g_{KNY} + \Delta_{KNY}(0) \ ,
\eqno (3)
$$
and strongly dependent on the nature and parameters of this breaking in
QCD$^{15}$.

\par Note that the most advanced analyses$^{16}$ still yield errors of order
unit on $g^2_{KNY} /{4\pi}$, {\sl one order of magnitude} larger than
errors in the $\pi N$ case! To reach definite conclusions, one needs thus one
order of magnitude improvements$^{17}$ on the determination of the couplings:
due to the cancellations
involved in the dispersive integrals, this requires even better
improvements in the actual data. For the $KN\Sigma$ coupling (the smaller and
more uncertain of the two) $K_S$ regeneration in hydrogen is the best source
of information, directly giving (through isospin symmetry) the
charge--odd combination of $K^\pm n$ amplitudes required for the
integrands, and thus eliminating part of the above--mentioned cancellations,
while at the same time dispensing us from the uncertainties related to the
neutron--amplitude extraction from $K^\pm d$ data$^{18}$.

\par It is also worthwhile mentioning that deuterium scattering data {\sl do
not exist at all} below a momentum of about $400 MeV/c$, and one has to
replace them with extrapolations (with inherently larger errors) over a quite
sizeable energy interval. Scattering data on deuterium are also interesting
{\sl per se}, as a testing ground of multichannel, three--body
calculations.

\par A second problem, related both to the nature of the chiral--symmetry
breaking and to the structure of the baryon octet wavefunctions, is the
determination of the so--called $\sigma$--terms, proportional to the
elastic, forward scattering amplitudes for zero four--momentum kaons.
Their values can be extracted in the most economical way using the values
for the amplitudes at zero laboratory--energy, at momentum transfers $t \leq
0$ obtainable from fixed--$t$ dispersion relations$^{19}$.

\par Amplitudes at zero laboratory--energy do not have any special theoretical
significance in themselves: it is only because the Cheng--Dashen point for
$KN$--scattering amplitudes, $t = 2m_K^2$, $s = u = M_N^2$, lies in the
$t$--channel unphysical region, that one is forced to construct a ``modified
Cheng--Dashen technique'', employing zero--energy amplitudes, to get at the
$KN$ $\sigma$--terms. These, and in particular their $I_t = 0$ part
$$
\sigma _{KN}^{(0)} = {{m_s+m}\over {m_s-m}} \ {{1+y}\over {1-y}} \ \lbrack
\langle N|\ -{3\over 4}\ H_8(0)\ |N \rangle \rbrack + O({{m_d - m_u}\over
{m_s - m}}) \ ,
\eqno (4)
$$
are an extremely sensitive and direct measurement of the scalar,
strange--quark density in the nucleon, measured by the ratio
$$
y = \langle N | \bar s s | N \rangle / \langle N | {1 \over 2} (\bar u u +
\bar d d) | N \rangle \ .
\eqno (5)
$$

\par The above $\sigma$--term is a much more direct
indication of the value of $y$ than the $\pi N$ one
$$
\sigma_{\pi N} = {m\over {m_s - m}} \ {1\over {1-y}} \ \lbrack \langle N|\ -3
\ H_8(0)\ |N \rangle \rbrack \ ,
\eqno (6)
$$
where it is hard to separate the strong dependence on $y$ from the equally
strong one on the light--quark mass ratio $m_s/m$.

\par The ratio $y$ has been the subject of a recent, lively theoretical
debate$^{20}$ (mostly based on model--dependent assumptions), with
wide--ranging implications, from OZI--rule violations in scattering and
production of hadrons, through the validity of lattice approaches to
baryon structure and spectroscopy, to the equation of state of dense baryonic
matter, with the enticing possibility of ``kaon condensation'' in primordial
baryogenesys, dense neutron stars, and relativistic heavy--ion
collisions$^{21}$.

\par Employing $t$--channel unitarity and analyticity, plus the hypothesis that
the $S$--wave dominates this channel at low values of $t$, the
on--mass--shell, zero--energy amplitudes have been connected to the
$\sigma$--terms, via the phases $\phi_I(t)$ of the isospin--$I$,
$t$--channel $\bar NN \rightarrow \bar KK$ amplitude, related to the
$S$--wave meson--meson ones via coupled--channel unitarity in an $N/D$
representation$^{19}$, and derived a relation of the form
$$
D^{(+)}_I(t) = {{2 \sigma_{KN}^{(I)}} \over f_K^2} {P_I(t)\over
P_I(2m_K^2)} {\exp
[{{t\over\pi}\int {{\phi_I(t^{\prime}) dt^{\prime}}\over{t^{\prime} - t}}}} ]
+ Born\ term\ contributions\ ,
\eqno (7)
$$
where $P_I(t)$ is a suitably smooth function of $t$, analytic in the
$t$--plane cut from $t_L < 0$ to $t = 4m_K^2$. By dividing the zero--energy
amplitudes (obtained from dispersion relations) by the Omn\`es function in eq.
(7), and subtracting the Born terms, one extrapolates this
representation to $t = 2m_K^2$ to extract the $I_t = 0$ $\sigma$--term,
and from it the ratio $y$.

\par In contrast with the small absolute uncertainty of the method when
applied to $\pi N$--amplitudes$^{19,22}$, its application to the $KN$ case
carries an {\sl intrinsic} uncertainty of the order of 100 MeV for both
$K^\pm N$ $\sigma$--terms, which could be adequate to discriminate
between at least some of the existing, model--dependent predictions$^{20}$,
were not the errors coming from the data several times larger$^{19,23}$,
and indeed comparable with the estimates themselves.
A definite answer to the $\sigma$--term problem,
{\sl both} for the $KN$ {\sl and} the $\pi N$ systems,
could thus come only through a drastic reduction in the errors of the first
one,
extremely sensitive to those on the low--energy and unphysical $\bar KN$
regions$^{23}$, due to large cancellations in the dispersive integrals and
between these latter and the Born terms.

\par Another problem, which for a long time has puzzled baryon
spectroscopists, is the nature of the $\Lambda(1405)$, shifting with
times from the $S$--matrix language (CDD pole, $\bar KN$ bound
state, or resonance?) to the quark--gluon one ($qqq$ baryon, $\bar KN$ bound
state, multiquark or hybrid state?). Indeed even its correct reproduction
remains an outstanding problem in quark--model phenomenology$^{24}$.

\par The state can be directly observed (and has indeed been seen$^{13}$)
only in $\Sigma\pi$ (and possibly in $\Lambda\pi\pi$) inelastic production:
the determination of its coupling to the channel
$\bar KN$ requires an {\sl extremely
stable} extrapolation of the parametrization for the threshold--region into
the unphysical one, possible only with much smaller error corridors and
more stringent analyticity constraints than those presently available$^{25}$.

\par The $\Lambda(1405)$ nature reflects on the probabilities for
radiative capture at threshold (the ``strange'' analogues of the Panofsky
ratio in the $\pi N$ system) as well: present measurements are just above
observability$^{26}$ (mostly because of the extremely high backgrounds due
to the pion contamination of the beams), and therefore unable to decide
between different, if not conflicting, theoretical expectations$^{27}$.

\par The high purity of the $\phi$--factory as a kaon source makes it an ideal
environment for this kind of physics.

\par Last, but not least, we wish to mention the still open mystery of kaonic
hydrogen: the three experiments performed up to now on this system$^{28}$ have
collected a total of only a few tens of events {\sl interpreted} as atomic
$K$--lines, above backgrounds orders of magnitude larger.
Even their identification as {\sl bona--fide} kaonic--hydrogen spectral lines
is open to questioning$^{29}$. Also, to the best of our knowledge,
all models purported to reproduce them conflict either with $\bar KN$
amplitude analyses or with simple, quantum--mechanical pictures of
absorptive processes: thus these models
either failed to solve this mystery, or
were able to explain the data only at the price of highly non--smooth or
even unphysical energy behaviors of the amplitudes$^{30}$.

\par An experiment, planned to clearly identify and measure energies and
widths of the $K^-p$ (and $K^-d$) $K$--lines,
could either eliminate the mystery or open the way to new and unforeseen
physics in this very--low--energy domain of the strong interactions.

\vglue 0.6truecm
\noindent{\bf 4. A SKETCH OF AN EXPERIMENTAL AND THEORETICAL PROGRAM FOR
$KN$ PHYSICS AT THE NEW MACHINES. }
\vglue 0.3truecm

\par This last section of our review will be covering what we believe are the
programs to be carried out, both by the experimentalists working
at the new machines
and by the theorists in order
to fully exploit the data these machines will be
producing.

\par In 1990, at the Vancouver ``Workshop on Science at the KAON Factory'',
we presented a list of experiments we then thought could, and ought to,
be carried on
{\sl only} at such a facility$^{31}$. The new possibilities, opening up at a
{\sl suitable} $\phi$--factory such as DA$\Phi$NE$^{32}$, force us to slightly
revise that list.

\par Its first item was ``a dedicated, low--energy ($p_L \leq$
300 MeV/c) $K^- p$ or $K^- d$'' scattering ``experiment with the cleanest
beam--line afforded by present--day technology'': probably for $p_L \leq$ 125
MeV/c a {\sl dedicated} detector at the $\phi$--factory DA$\Phi$NE
can win over a fixed--target experiment, due to the
{\sl extreme} cleanliness of the former in this momentum range,
even despite its low kaon--production rate. A possibility we did
not mention was that of doing experiments with polarized targets:
fixed--target machines are indeed the {\sl only} place where these
measurements can be performed. The latter, essential to any good
amplitude analysis,
have been up to now unavailable for momenta below
about 800 MeV/c. Note in this respect that DA$\Phi$NE can produce about
$3 \times 10^6$ $K^- p$ interactions in a typical apparatus$^{32}$ in a
``Snowmass year'' ({\sl i.e.} $10^7$ s) of experimentation, enough to
measure angular distributions in all channels, and also polarizations, but
the latter only for the self--analyzing, final--hyperon states.

\par Second item on that list was a ``good quality $K^0_S$--regeneration
experiment on hydrogen (and deuterium) with an intense, low--momentum $K^0_L$
beam'': here DA$\Phi$NE can help at very low momenta (110
MeV/c only), measuring {\sl tagged} $K^0_S$ regenerations
(together with {\sl all
inelastic channels}) in the same apparatus used to measure $K^-$ (and $K^+$)
interactions with hydrogen (or deuterium, changing the gas filling the
``target'' fiducial volume), again with ``yearly'' rates of the order of
$10^6$ events.

\par The third item was ``a kaonic--hydrogen and deuterium experiment'':
Shimoda should see presented the first results coming out of the new japanese
experiment, which hopefully should solve (or reopen?) the mystery. Undeniably,
KAON fluxes shall make statistics even better, and thus offer a further,
stringent constraint on $KN$ amplitude analyses, fixing the $K^-p$ and $K^-d$
scattering lengths via the energy shifts and widths of the atomic $1s$ levels.

\par Last item on the list was ``an elastic--scattering experiment in the
Coulomb--nuclear interference region to map $\rho = D/A$'', the
real--over--imaginary--part ratio, ``over reasonably small energy steps'',
where of course the wide momentum ranges that will be provided by KAON should
prove invaluable.

\par We wish to add now to this list the remark that a DA$\Phi$NE detector
dedicated to kaon experiments on gaseous $H_2$ and $D_2$ can continue is
active life, without substantial changes, to measure $K^+$--, $K^-$--, and
$K^0_L$--interactions on heavier gases as well ($He$, $N_2$, $O_2$, $Ne$,
$Ar$, $Kr$, $Xe$), exploring not only the properly ``nuclear'' aspects
of these interactions, such as the aforementioned nucleon ``swelling'' in
nuclei$^8$, but also producing $\pi\Sigma$, $\pi\Lambda$ and $\pi\pi\Lambda$
systems with invariant masses in the $\bar KN$ unphysical region, with
statistics substantially higher than those now available$^{13}$, due to
the $\simeq 4\pi$ geometry allowed by a colliding--beam--machine
detector.

\par All these experiments will provide data of the same quantity and quality
now available only for the $\pi N$ system: theoretical tools for their
analysis must thus be improved as well, to meet the standards required by
this, long awaited for, ``forward leap'' in the quality of the $KN$ data.
Tools of just this level have since long been provided for $\pi N$ amplitude
analysis by the so--called ``Karlsruhe--Helsinki collaboration'' headed over
the years by Prof. G. H\"ohler$^{33}$; their software can not be
straightforwardly ``imported'' to do $KN$ amplitude analyses,
mainly because of
the complicated analytic structures of the low-energy $\bar KN$
strong--interaction amplitudes. Much for the same reason, the dispersive
treatment of Coulomb corrections developed at NORDITA$^{34}$ can not be
immediately transferred to the strange sector. It has to be recalled that
the old data were {\sl almost always} analyzed using,
for these corrections, the old, approximate formula
of Dalitz and Tuan$^{35}$, which may be inapplicable to a strongly absorptive
interaction close to threshold$^{29}$.

\par Since the basic principles on which both approaches are based have to hold
also for the $KN$ system, as for the $\pi N$ one, it remains for us
to work out the details of a partial--wave--analysis procedure,
applicable to a system strongly absorptive at threshold such as the $\bar KN$
one, and possessing {\sl ab--initio} the following requirements: i)
consistency with both fixed--$t$ and partial--wave dispersion relations; ii)
crossing symmetry (and isotopic--spin symmetry as well, to describe {\sl
simultaneously} charge--exchange and regeneration data); iii) analyticity in
$t$ beyond the Lehman ellipses, with the {\sl correct} low--mass,
$t$--channel--cut discontinuities given by the $\pi\pi$ cut; iv) a {\sl
complete} treatment of Coulomb corrections.

\par Both of us have in the past carried out parts of this program (as many
others have also done more or less at the same epoch),
but only for {\sl limited purposes}, such as extrapolations
either to the hyperon poles$^{16}$ or to the Cheng--Dashen point$^{19}$, or
studies of Coulomb effects at threshold$^{29}$: what remains to be done is a
merging together of all these techniques into a ``global'' analysis, on which
work is presently under way.

\par In this perspective we have advanced the proposal to I.N.F.N. for a
program of extensive collaboration, code--named KILN (for ``Kaon Interactions
at Low energies with Nucleons''), which has already received an initial,
and thus limited, financial support.

\par Participation in this collaboration is highly welcome, and we take this
occasion for calling upon all theorists wich have been or wish to be active
in this still open and very much alive (despite greatly exaggerated rumors on
the contrary$^{36}$) field of particle physics.

\vglue 0.6truecm
\centerline{\bf ACKNOWLEDGEMENTS }
\vglue 0.3truecm

\par One of us (G. V.) would like to express here his gratitude to
Colciencias for their substantial support to his participation in this
Conference. Also, exchanges of information with theorists and
experimentalists involved in the planning of experiments at both
machines, KAON and DA$\Phi$NE, is gratefully acknowledged by both authors.

\vglue 1.5truecm
\centerline{\bf REFERENCES }
\vglue 0.3truecm

\item{1.} G. Violini, in: {\sl ``Low and Intermediate Energy Kaon-Nucleon
Physics''}, ed. by E. Ferrari and G. Violini (D. Reidel, Dordrecht 1981),
p. 419.
\item{2.} G. Vignola, in: {\sl ``Workshop on Physics and Detectors for
DA$\Phi$NE''}, ed. by G. Pancheri (I.N.F.N., Frascati 1991), p. 11.
\item{3.} M. K. Craddock, in: {\sl ``Workshop on Science at the KAON
Factory''}, ed. by D. R. Gill (TRIUMF, Vancouver 1991), Vol. I, p. 7.
\item{4.} J. Beveridge, in: {\sl ``Workshop on Science at the KAON
Factory''}, ed. by D. R. Gill (TRIUMF, Vancouver 1991), Vol. I, p. 19.
\item{5.} G. Alexander, {\sl et al.}, {\sl. Phys. Lett.} {\bf B 58} (1975) 484.
\item{6.} G.W. London, paper 135 presented at the
{\sl ``XVIIth Conference on High Energy Physics''} (London, 1974).
\item{7.} G. C. Oades, in: {\sl ``Low and Intermediate Energy Kaon-Nucleon
Physics''}, ed. by E. Ferrari and G. Violini (D. Reidel, Dordrecht 1981),
p. 53.
\item{8.} E. Piasetzky, {\sl Nuovo Cimento} {\bf 102 A} (1989) 281.
\item{9.} N. M. Queen and G. Violini, {\sl ``Dispersion Theory in High Energy
Physics''} (McMillan, London 1974).
\item{10.} N. M. Queen, M. Restignoli and G. Violini, {\sl Fortschr. Phys.}
{\bf 17} (1969) 467; {\sl ibid.} {\bf 21} (1973) 569.
\item{11.} J. K. Kim, {\sl Phys. Rev. Lett.} {\bf 19} (1967) 1074; B. R.
Martin and M. Sakitt, {\sl Phys. Rev.} {\bf 183} (1969) 1345; A. D. Martin
and G. G. Ross, {\sl Nucl. Phys.} {\bf B 16} (1970) 479.
\item{12.} A. D. Martin, in: {\sl ``Low and Intermediate Energy Kaon-Nucleon
Physics''}, ed. by E. Ferrari and G. Violini (D. Reidel, Dordrecht 1981), p.
97; {\sl Nucl. Phys.} {\bf B 179} (1981) 33; R. H. Dalitz and A. Deloff, {\sl
``The Analysis of Low--Energy $KN$ Reaction Data''} (1991, to be published).
\item{13.} R. J. Hemingway, {\sl Nucl. Phys.} {\bf B 253} (1985) 742.
\item{14.} C. A. Dom\'\i nguez, {\sl Riv. Nuovo Cimento} {\bf 8} (1985) N. 6.
\item{15.} H. F. Jones and M. D. Scadron, {\sl Phys. Rev.} {\bf D 11} (1975)
174; N. H. Fuchs, H. Sazdjian and J. Stern, {\sl Phys. Lett.} {\bf B 238}
(1990) 381.
\item{16.} G. K. Atkin, B. Di Claudio, G. Violini and N. M. Queen, {\sl
Phys. Lett.} {\bf B 95} (1980) 447, and in: {\sl ``Low and Intermediate Energy
Kaon-Nucleon Physics''}, ed. by E. Ferrari and G. Violini (D. Reidel,
Dordrecht 1981), p. 131; J. Antol\'\i n, {\sl Phys. Rev. D} {\bf 43} (1991)
1532.
\item{17.} P. M. Gensini, {\sl J. Phys.} {\bf G 7} (1981) 1315.
\item{18.} M. Lusignoli, M. Restignoli and G. Violini, {\sl Phys. Lett.} {\bf
B 24} (1967) 296.
\item{19.} P. M. Gensini, {\sl J. Phys. G} {\bf 7} (1981) 1177; {\sl Nuovo
Cimento} {\bf 84 A} (1984) 203.
\item{20.} R. L. Jaffe and C. L. Korpa, {\sl Comments Nucl. Part. Phys.}
{\bf 17} (1987) 163; R. L. Jaffe, {\sl Nucl. Phys.} {\bf A 478} (1988) 3c;
P. M. Gensini, {\sl Nuovo Cimento} {\bf 102 A} (1989) 75, erratum 1181. See
also B. L. Ioffe and M. Karliner, {\sl Phys. Lett.} {\bf B 247} (1990) 387; G.
Cl\'ement, M. D. Scadron and J. Stern, {\sl J. Phys.} {\bf G 17} (1991) 199.
\item{21.} D. B. Kaplan and A. E. Nelson, {\sl Phys. Lett.} {\bf B 175} (1986)
57, {\sl erratum} {\bf B 179} (1986) 409; {\sl Nucl. Phys.} {\bf A 479} (1988)
273c; A. E. Nelson and D. B. Kaplan, {\sl Phys. Lett.} {\bf B 192} (1987) 193;
{\sl Nucl. Phys.} {\bf A 479} (1988) 285c.
\item{22.} J. Gasser, H Leutwyler and M. E. Sainio, {\sl Phys. Lett.} {\bf B
253} (1991) 252; see also the talks presented by J. Gasser and M. E. Sainio at
the {\sl ``IVth Int. Symposium on Pion-Nucleon Physics and the Structure of
the Nucleon''}, to be published in {\sl $pi N$ Newsletter} {\bf 4} (1991).
\item{23.} B. Di Claudio, G. Violini and A. M. Rodr\'\i guez--Vargas, {\sl
Lett. Nuovo Cimento} {\bf 26} (1979) 555; B. Di Claudio, A. M. Rodr\'\i
guez--Vargas and G. Violini, {\sl Z. Phys.} {\bf C 3} (1979) 75; A. M.
Rodr\'\i guez--Vargas and G. Violini, {\sl Z. Phys.} {\bf C 4} (1980) 135;
A. D. Martin and G. Violini, {\sl Lett. Nuovo Cimento} {\bf 30} (1981) 105;
A. M. Rodr\'\i guez--Vargas, in: {\sl ``Low and Intermediate Energy
Kaon-Nucleon Physics''}, ed. by E. Ferrari and G. Violini (D. Reidel,
Dordrecht 1981), p. 331.
\item{24.} W. Lucha, E. F. Sch\"oberl and D. Gromes, {\sl Phys. Rep.}
{\bf 200} (1991) 127.
\item{25.} R. H. Dalitz and J. G. McGinley, in: {\sl ``Low and Intermediate
Energy Kaon-Nucleon Physics''}, ed. by E. Ferrari and G. Violini (D. Reidel,
Dordrecht 1981), p. 381; R. H. Dalitz, J. G. McGinley, C. Belyea and S.
Anthony, in : {\sl ``Int. Conf. on Hypernuclear and Kaon Physics''}, ed. by B.
Povh, {\sl report MPI--H--1982--V20} (MPI, Heidelberg 1982), p. 201;
R. H. Dalitz and A. Deloff, {\sl J. Phys.} {\bf G 17} (1991) 289.
\item{26.} D. Horvath, {\it et al.}, {\sl ``3rd Int. Sympos. on Pion--Nucleon
and Nucleon--Nucleon Physics''}, ed. by S. P. Kruglov, {\it et al.} (IYaF,
Leningrad 1989), Vol. 1, p. 375; B. L. Roberts, {\it et al.}, {\sl Nuovo
Cimento} {\bf 102 A} (1989) 145; D. A. Whitehouse, {\it et al.}, {\sl Phys.
Rev. Lett.} {\bf 63} (1989) 1352.
\item{27.} R. L. Workman and H. W. Fearing, {\sl Phys. Rev.} {\bf D 37} (1988)
3117; Y.--S. Zhong, A. W. Thomas, B. K. Jennings and R. C. Barrett, {\sl Phys.
Rev.} {\bf D 38} (1989) 837; J. Lowe, {\sl Nuovo Cimento} {\bf 102 A} (1989)
167; R. Williams, C.--R. Ji and S. R. Cotanch, {\sl Phys. Rev.} {\bf D 41}
(1990) 1449; H. Burkhardt and J. Lowe, {\sl Phys. Rev.} {\bf C 44} (1991) 607.
\item{28.} J. D. Davies, {\sl et al., Phys. Lett.} {\bf B 83} (1979) 55; M.
Izycki, {\sl et al., Z. Phys.} {\bf A 297} (1980) 11; P. M. Bird, {\sl et al.,
Nucl. Phys.} {\bf A 404} (1983) 482.
\item{29.} P. M. Gensini and G. R. Semeraro, in: {\sl ``Perspectives on
Theoretical Nuclear Physics''}, ed. by L. Bracci, {\sl et al.} (ETS Ed., Pisa
1986), p. 91; C. J. Batty, in: {\sl ``First Workshop on Intense Hadron
Facilities and Antiproton Physics''}, ed. by T. Bressani, F. Iazzi and G.
Pauli (S.I.F., Bologna 1990), p. 117.
\item{30.} R. C. Barrett, {\sl Nuovo Cimento} {\bf 102 A} (1989) 179; C. J.
Batty and A. Gal, {\sl Nuovo Cimento} {\bf 102 A} (1989) 255.
\item{31.} P. M. Gensini and G. Violini, {\sl ``Workshop on Science at the
KAON Factory''}, ed. by D. R. Gill (TRIUMF, Vancouver 1991), Vol. II, p. 193.
\item{32.} P. M. Gensini, in {\sl ``Workshop on Physics and Detectors for
DA$\Phi$NE''}, ed. by G. Pancheri (INFN, Frascati 1991), p. 453.
\item{33.} G. H\"ohler, F. Kaiser, R. Koch and E. Petarinen, {\sl ``Handbook
of Pion-Nucleon Scattering'', Physik Daten Nr. 12-1} (Fachinformationszentrum,
Karlsruhe 1979); R. Koch, in: {\sl ``Low and Intermediate Energy
Kaon-Nucleon Physics''}, ed. by E. Ferrari and G. Violini (D. Reidel,
Dordrecht 1981), p. 1; see also the update by G. H\"ohler, {\sl $\pi N$
Newsletter} {\bf 2} (1990) 1.
\item{34.} B. Tromborg, S. Waldenstr\"om and I. \O verb\o, {\sl Phys. Rev.}
{\bf D 15} (1977) 725; {\sl Helv. Phys. Acta} {\bf 51} (1978) 584.
\item{35.} R. H. Dalitz and S. F. Tuan, {\sl Ann. Phys. (N.Y.)} {\bf 10}
(1960) 307.
\item{36.} The same also happened to Mark Twain, from which we took the
liberty of borrowing the pun.

\bye